\documentstyle[12pt,epsfig,amssymb]{article}

\def\T{\textstyle}

\def\SS{\scriptscriptstyle}

\def\lc{\Lambda_c}
\def\CP{C\hspace{-0.7mm}P}

\def\NF{N\hspace{-0.35mm}F}

\title{
\vspace*{-1.5cm}
\begin{flushright}
{\normalsize DO--TH 98/11\\[1.2cm]}
\end{flushright}
{\Large \bf A New Analysis of \boldmath $B_{6}^{(1/2)},$ \unboldmath
\boldmath $B_{8}^{(3/2)},$ \unboldmath and the \boldmath $\Delta I=1/2$
\unboldmath Rule in the \boldmath $1/N_c$ \unboldmath 
Expansion for \boldmath $K\rightarrow \pi\pi$ \unboldmath Decays  
\hspace{-3mm}
}
\thanks{Talk presented at the XVI Autumn School and Workshop on Fermion
Masses, Mixing and CP Violation, Lisboa, Portugal, 6-15 October 1997.}
\vspace*{0.3cm}
}
\vspace{-1.3cm}
\author{ \vspace*{0.0cm}\\
\noindent
\large Gerhard O. K\"ohler\\
\normalsize {\it Institut f\"ur Physik, Universit\"at Dortmund,
D-44221 Dortmund, Germany} \\[0.5cm]
}
\date{}
\begin{document}
\maketitle
\thispagestyle{empty}
\vspace*{-1.0cm}
\begin{abstract}
\footnotesize
We present a new calculation of long-distance contributions to the
$K\rightarrow\pi\pi$ amplitudes using the $1/N_c$ expansion within the 
framework of chiral perturbation theory. Along these lines we compute the 
chiral loop corrections to the operators $Q_1$, $Q_2$, $Q_6$, and $Q_8$. 
In the numerical analysis we present values for $B_6^{(1/2)}$ and
$B_8^{(3/2)}$. We discuss also the implications for a theoretical 
understanding of the $\Delta I=1/2$ selection rule.  
\end{abstract}
\normalsize
%
\section{Introduction}

Within the standard model the calculation of the $K\rightarrow \pi\pi$
decay amplitudes is based on the effective low-energy hamiltonian for
$\Delta S=\nolinebreak 1$ transitions \cite{delS},
\begin{equation}
{\cal H}_{ef\hspace{-0.5mm}f}^{\SS \Delta S=1}=\frac{G_F}{\sqrt{2}}
\;\xi_u\sum_{i=1}^8 c_i(\mu)Q_i(\mu)\hspace{1cm} (\,\mu<m_c\,)\;,
\end{equation}
\begin{equation}
c_i(\mu)=z_i(\mu)+\tau y_i(\mu)\;,\hspace*{1cm}\tau=-\xi_t/\xi_u\;,
\hspace*{1cm}\xi_q=V_{qs}^*V_{qd}^{}\;,
\end{equation}
where the Wilson coefficient functions $c_i(\mu)$ of the local four-fermion
operators $Q_i(\mu)$ are obtained using perturbative renormalization group
techniques \cite{BJL,CFMR}. $Q_1$-$Q_5$ and $Q_7$ are composed of products of 
two currents, $Q_6$ and $Q_8$ are density-density operators. In view of 
phenomenological applications $Q_1$ and $Q_2$, as well as, $Q_6$ and $Q_8$ 
are of special importance, where
\begin{equation}
Q_1=4\bar{s}_L\gamma^\mu d_L\,\bar{u}_L\gamma_\mu u_L\,,\hspace{1cm} 
Q_2=4\bar{s}_L\gamma^\mu u_L\,\bar{u}_L\gamma_\mu d_L\,,\hspace{1cm} 
\end{equation}
\begin{equation}
Q_6=-8\sum_{q=u,d,s}\bar{s}_Lq_R\,\bar{q}_Rd_L\,,\hspace{1cm} 
Q_8=-12\sum_{q=u,d,s}e_q\bar{s}_Lq_R\,\bar{q}_Rd_L\,,\hspace{1cm} 
 \end{equation}
with $q_{L,R}=\frac{1}{2}(1\pm\gamma_5)q$ and $e_q=(2/3,\,-1/3,\,-1/3)$. 
The hadronic matrix elements of the operators, 
$\langle Q_i(\mu)\rangle_I\equiv\langle\pi\pi,I|Q_i(\mu)|K^0\rangle$,
are difficult to calculate but can be estimated using non-perturbative
techniques generally for $\mu$ around a scale of $1\,$GeV. In the following
analysis we will use the $1/N_c$ expansion within the framework of the 
chiral effective lagrangian for pseudoscalar mesons. 

The calculation of chiral loop effects motivated by the $1/N_c$ expansion 
was introduced in Ref.~\cite{BBG} to investigate the $\Delta I=1/2$ selection 
rule. Experimentally one finds $Re\,A_0/Re\,A_2\simeq 22.2$ for the ratio of
the isospin amplitudes. Keeping only the dominant terms the theoretical 
expression for this ratio reads 
\begin{equation}
\frac{Re\,A_0}{Re\,A_2}\,\simeq\,\frac{z_1(\mu)\langle Q_1(\mu)\rangle_0
+z_2(\mu)\langle Q_2(\mu)\rangle_0+z_6(\mu)\langle Q_6(\mu)\rangle_0}
{z_1(\mu)\langle Q_1(\mu)\rangle_2+z_2(\mu)\langle Q_2(\mu)\rangle_2}\,.
\end{equation}

It was recognized, long ago, that the vacuum saturation approximation (VSA)
\cite{VSA} completely fails in explaining the $\Delta I=1/2$ rule. The
authors of Ref.~\cite{BBG} considered chiral loop corrections to the
current-current operators, $Q_1$ and $Q_2$, and included the gluon penguin 
operator $Q_6$ at the tree level. As a net result they obtained a large 
enhancement of the $\Delta I=1/2$ decay amplitude which, however, was found 
to be still below the experimental value ($Re\,A_0=3.33\times 10^{-7}\,$GeV). 
One might note that the agreement with experiment is not improved by
including the NLO values for the $z_i$ \cite{BBL}.

The $\Delta I=1/2$ selection rule is governed by $Q_1$ and $Q_2$, while
the ratio $\varepsilon'/\varepsilon$ which measures the direct $\CP$ 
violation in $K\rightarrow\pi\pi$ decays is dominated by the density-density 
operators $Q_6$ and $Q_8$, that is to say, 
approximately\footnote{Recent reviews on $\CP$ violation in
$K\rightarrow\pi\pi$ decays can be found in Ref.~\cite{cprev}.} 
\begin{equation}
\varepsilon'/\varepsilon\,\propto\, y_6\langle Q_6\rangle_0-y_8
\langle Q_8\rangle_2\,+\, \mbox{`$\pi$-$\eta$-$\eta^{\prime}$ mixing'}\,.
\end{equation}
It is of special importance to investigate whether the $1/N_c$ (chiral loop)
corrections significantly affect the large cancellation between the gluon and 
the electroweak penguin contributions obtained at the tree level in 
Ref.~\cite{Buch}. 

In this talk we present a new analysis of the hadronic matrix elements 
of the operators $Q_1$, $Q_2$, $Q_6$, and $Q_8$ in which we include the 
first order corrections in the twofold expansion in powers of external 
momenta, $p$, and the ratio $1/N_c$. Our main improvement concerns the
matching of short- and long-distance contributions to the decay amplitude, 
by adopting a modified identification of virtual momenta in the chiral loops. 
In Section 2 we briefly review the framework of the effective low-energy
approach and discuss the matching of short- and long-distance contributions. 
In Section 3 we list the analytic expressions for the hadronic matrix 
elements. Finally, in Sections 4 and 5 we present our numerical results.

We will focus here on the computation of the $K\rightarrow\pi\pi$ matrix
elements. A more detailed discussion of the general method and the operator
evolution at long distances is presented by T. Hambye and P. Soldan in these
proceedings, and  some additional details can be found in 
Refs.~\cite{HKPSB,TH}. 

\section{General Framework}
We start from the chiral effective lagrangian for pseudoscalar mesons which 
involves an expansion in momenta where terms up to $O(p^4)$ are included
\cite{GaL}. Keeping only terms which contribute to the $K\rightarrow
\pi\pi$ matrix elements and are leading in $N_c$ it reads
\begin{eqnarray}
{\cal L}_{ef\hspace{-0.5mm}f}&=&\frac{f^2}{4}\Big(
\langle D_\mu U^\dagger D^{\mu}U\rangle
+\frac{\alpha}{4N_c}\langle \ln U^\dagger -\ln U\rangle^2
+r\langle {\cal M} U^\dagger+U{\cal M}^\dagger\rangle\Big) 
+r^2 H_2 \langle {\cal M}^\dagger{\cal M}\rangle \nonumber\\[1mm] 
&& +rL_5\langle D_\mu U^\dagger D^\mu U({\cal M}^\dagger U
+U^\dagger{\cal M})\rangle+rL_8\langle {\cal M}^\dagger U{\cal M}^\dagger U
+{\cal M} U^\dagger{\cal M} U^\dagger \rangle \label{lagr}\,,
\end{eqnarray}
with $D_\mu U=\partial_\mu U-ir_\mu+iUl_\mu$, $\langle A\rangle$ denoting 
the trace of $A$ and $M = diag( m_u,m_d,m_s)$. $f$ and $r$ are free 
parameters related to the pion decay constant $F_\pi$ and to the quark 
condensate, respectively, with $r = - 2 \langle \bar{q}q\rangle/f^2$.
The complex matrix $U$ is a non-linear representation of the pseudoscalar
meson nonet:
\begin{equation}
U=\exp\frac{i}{f}\Pi\,,\hspace{1cm} \Pi=\pi^a\lambda_a\,,\hspace{1cm} 
\langle\lambda_a\lambda_b\rangle=2\delta_{ab}\,. 
\end{equation}
In terms of the physical states
\begin{equation}
\Pi=\left(
\begin{array}{ccc}
\T\pi^0+\frac{1}{\sqrt{3}}a\eta+\sqrt{\frac{2}{3}}b\eta'
& \sqrt2\pi^+ & \sqrt2 K^+  \\[2mm]
\sqrt2 \pi^- & \T
-\pi^0+\frac{1}{\sqrt{3}}a\eta+\sqrt{\frac{2}{3}}b\eta' & \sqrt2 K^0 \\[2mm]
\sqrt2 K^- & \sqrt2 \bar{K}^0 & 
\T -\frac{2}{\sqrt{3}}b\eta+\sqrt{\frac{2}{3}}a\eta'
\end{array} \right)\,,
\end{equation}
with
\begin{equation}
a= \cos \theta-\sqrt{2}\sin\theta\,, \hspace{1cm}
\sqrt{2}b=\sin\theta+\sqrt{2}\cos\theta\,,
\end{equation}
and the $\eta-\eta'$ mixing angle $\theta$ for which we take 
$\theta=-19^\circ$ \cite{eta}. $\alpha$ is the instanton parameter which
reflects the explicit breaking of the $U_A(1)$. The symmetry violating term
must be included when treating the singlet $\eta_0$ as a dynamical degree of
freedom. 

The bosonic representations of the quark currents and densities are defined 
in terms of (functional) derivatives of the chiral action and the lagrangian, 
respectively:
\begin{eqnarray}
\bar{q}_{iL}\gamma^\mu q_{jL}&\hspace*{-1mm}\equiv& 
\hspace*{-1mm}\frac{\delta S}{\delta(l_\mu(x))_{ij}}\,=\,
-i\frac{f^2}{2}\big(U^\dagger\partial^\mu U\big)_{ji} \nonumber\\[1mm]
&&\hspace*{-1mm} 
+irL_5\big(\partial^\mu U^\dagger{\cal M}-{\cal M}^\dagger\partial^\mu U
+\partial^\mu U^\dagger U{\cal M}^\dagger U
-U^\dagger{\cal M} U^\dagger\partial^\mu U\big)_{ji}\,,\label{curr}\\[1mm]
\bar{q}_{iR}q_{jL}&\hspace*{-1mm}\equiv&\hspace*{-1mm}
-\frac{\delta{\cal L}_{ef\hspace{-0.5mm}f}}{\delta{\cal M}_{ij}}
\,=\,-r\Big(\frac{f^2}{4}U^\dagger+L_5\partial_\mu U^\dagger
\partial^\mu U U^\dagger +2rL_8U^\dagger{\cal M} U^\dagger
+rH_2{\cal M}^\dagger\Big)_{ji}\,,\hspace*{8.5mm} \label{dens}
\end{eqnarray}
and the right-handed expressions are obtained by parity transformation.
Eqs.~(\ref{curr}) and (\ref{dens}) allow us to express the operators $Q_i$
in terms of the pseudoscalar meson fields.

The $1/N_c$ corrections to the matrix elements $\langle Q_i\rangle_I$ are
calculated by chiral loop diagrams. The diagrams are ultraviolet divergent 
and are regularized by a finite cutoff. This procedure, which was introduced 
in Ref.~\cite{BBG}, is necessary in order to restrict the chiral 
lagrangian to the low-energy domain. Since we truncate the effective theory 
to pseudoscalar mesons, the cutoff has to be taken at or, preferably, below 
the ${\cal O} (1\,\,\mbox{GeV})$. 

To calculate the amplitudes we follow the lines of Ref.~\cite{BBG} and
identify the ultraviolet cutoff of the long-distance terms with the 
short-distance renormalization scale $\mu$. In carrying out this matching 
we pay special attention to the definition of the momenta inside the loop. 
This question must be addressed because the loop integrals, within the 
cutoff regularization, are not momentum translation invariant.

In the existing studies of the hadronic matrix elements the color singlet
boson connecting the two densities or currents was integrated out from the
beginning \cite{BBG,Buch,JMS1,EAP2}. Thus the integration variable was taken
to be the momentum of the meson in the loop, and the cutoff was the upper
limit of its momentum. As there is no corresponding quantity in the
short-distance part, in this treatment of the integrals there is no 
clear matching with QCD. The ambiguity is removed, for non-factorizable 
diagrams, by considering the two densities to be connected to each other 
through the exchange of the color singlet boson, as was already discussed 
in Refs.~\cite{TH,BB,FG,BGK,PS}. A consistent matching is then obtained by 
assigning the same momentum to the color singlet boson at long and short 
distances and by identifying this momentum with the loop integration 
variable (see Fig.~1). The effect of this procedure is to modify the loop 
integrals, which introduces noticeable effects in the final results. More 
important it provides an unambiguous matching of the $1/N_c$ expansion in 
terms of mesons to the QCD expansion in terms of quarks and gluons.\\[12pt] 
\noindent
\centerline{\epsfig{file=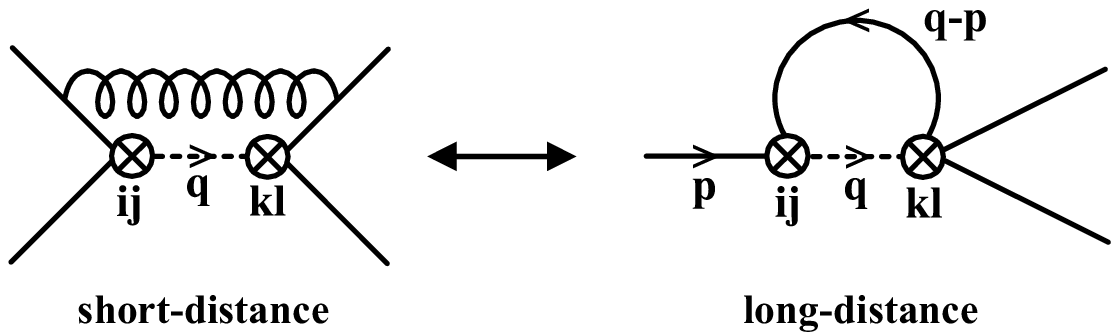,width=8.5cm}}\\[12pt]
\footnotesize
\centerline{Fig.\ 1. Matching of short- and long-distance contributions.}
\\[4pt]
\normalsize

The approach followed here leads to an explicit classification of the 
diagrams into factorizable and non-factorizable: there is no way to
establish a unique momentum prescription for the factorizable terms.
However, factorizable loop diagrams refer to the strong sector of the 
theory and give corrections whose scale dependence is absorbed in the 
renormalization of the chiral effective lagrangian. This statement is 
obvious in the case of the (conserved) currents and was proved explicitly, 
within the cutoff formalism, in the case of the bosonized 
densities.\footnote{For details see Ref.~\cite{HKPSB} and 
the contributions of T. Hambye and P. Soldan to these proceedings.}
Consequently, in the factorizable sector the chiral loop corrections do 
not induce ultraviolet divergent terms, i.e., the only remaining ultraviolet 
structure of the matrix elements is contained in the overall factor 
$r^2\sim 1/m_s^2$, whose scale dependence exactly cancels the factorizable 
evolution of the density-density operators $Q_6$ and $Q_8$ at short distances. 
The finite terms arising from the $1/N_c$ corrections to these operators, on 
the other hand, will not be absorbed completely in the renormalization of the 
various parameters. Nevertheless, as the complete sum of the factorizable 
diagrams is finite, we are allowed to send the scale of the {\it factorizable} 
terms to arbitrary high values, i.e., we are allowed to calculate the 
remaining finite corrections in dimensional regularization, which is 
momentum translation invariant and yields an unambiguous result for the
factorizable matrix elements. Finally, the non-factorizable loop diagrams 
have to be matched to the Wilson coefficients and should cancel scale 
dependences which arise from the (non-factorizable) short-distance expansion.
%
\section{Hadronic Matrix Elements}
Expanding the lagrangian of Eq.~(\ref{lagr}), as well as, the densities and 
currents of Eqs.~(\ref{curr}) and (\ref{dens}) in terms of the pseudoscalar 
fields, the various factorizable and non-factoriz\-able contributions to the 
$K\rightarrow \pi\pi$ matrix elements can be calculated from the diagrams 
of Figs.~2 and 3, respectively.\\[12pt] 
\noindent
\centerline{\epsfig{file=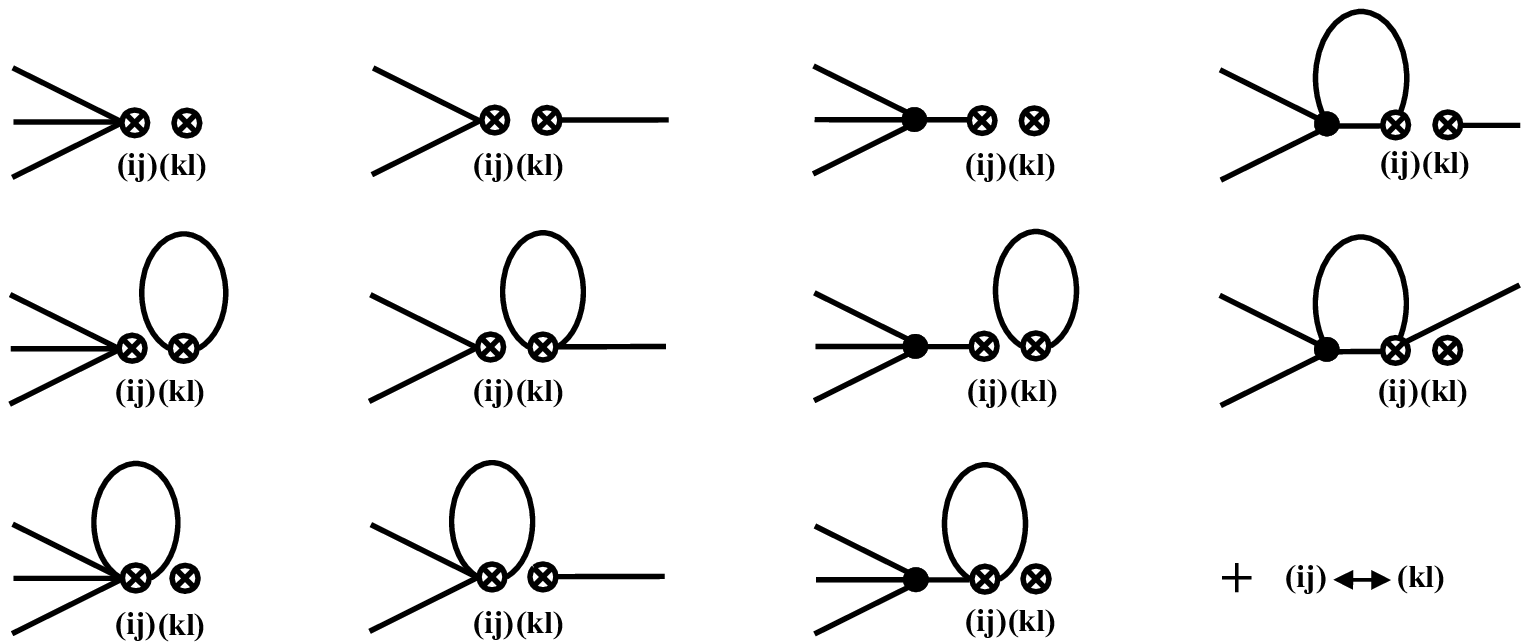,width=12.5cm}}\\[10pt]
\footnotesize
Fig.~2. Factorizable diagrams for the matrix elements of the operators
$Q_i$. Crossed circles represent the currents or densities, black circles
denote strong vertices. The lines represent the pseudoscalar mesons.
\\[12pt]
\normalsize

\noindent
\centerline{\epsfig{file=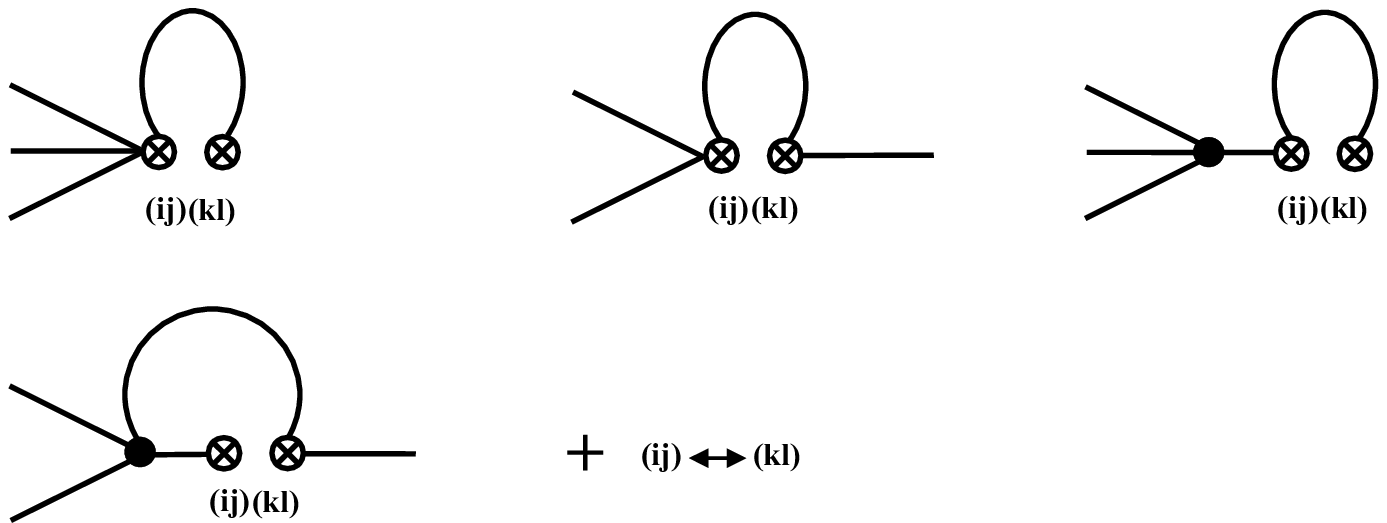,width=9.458cm}}\\[10pt]
\footnotesize
\centerline{Fig.~3. Non-factorizable diagrams for the matrix elements of 
the operators $Q_i$.} 
\\[4pt]
\normalsize

In addition, the wave function renormalization must be included, and the
values of the various parameters in the lagrangian must be expressed in
terms of physical quantities. This was done using dimensional regularization
as discussed in the previous section. We verified that the scale dependence
arising from the factorizable diagrams is completely canceled by the scale
dependence of the various parameters in the tree level expression of the
matrix elements. Finally, we obtain the following expressions:
\begin{eqnarray}
i\langle\pi^+\pi^-|Q_1|K^0 \rangle^F&=&
i\langle\hspace{0.5mm}\pi^0\hspace{0.5mm}\pi^0\hspace{0.5mm}|Q_2|
K^0\rangle^F\hspace{5.6mm}=\,\,\,0\,,\\[1mm]
i\langle\hspace{0.5mm}\pi^0\hspace{0.5mm}\pi^0\hspace{0.5mm}|Q_1|K^0
\rangle^F&=&-i\langle\pi^+\pi^-|Q_2|K^0 \rangle^F\,\,\,=\,\,\,
-X\left(1+\frac{4L_5^r}{F_\pi^2} m_\pi^2\right)\,,\label{fa1}\\[1mm]
i\langle\hspace{0.5mm}\pi^0\hspace{0.5mm}\pi^0\hspace{0.5mm}|Q_6|K^0 
\rangle^F&=&i\langle\pi^+\pi^-|Q_6|K^0\rangle^F\hspace{5.7mm}=\,\,\,-X
\frac{4L_5^r}{F_\pi^2}\left(\frac{2m_K^2}{{\hat m}+m_s}\right)^2 \,,\\
i\langle\hspace{0.5mm}\pi^0\hspace{0.5mm}\pi^0\hspace{0.5mm}|Q_8|K^0 
\rangle^F&=&X\frac{2L_5^r}{F_\pi^2}\left(\frac{2m_K^2}{\hat{m}+m_s}\right)^2\,,
\\[1mm]
i\langle\pi^+\pi^-|Q_8|K^0\rangle^F&=&
\frac{3}{4}\sqrt{2}F_\pi\left(\frac{2m_K^2}{\hat{m}+m_s}\right)^2
\left[1+\frac{4}{3F_\pi^2}(8m_K^2-11m_\pi^2)L_5^r\right.\nonumber
\hspace*{5mm}\\[1mm]
&&\left.-\frac{16}{F_\pi^2}(m_K^2-2m_\pi^2)L_8^r\right]\,.
\end{eqnarray}
with $X=\sqrt{2} F_\pi (m_K^2-m_\pi^2)$ and the couplings $L_5^r$, $L_8^r$ 
defined through the relations
\begin{eqnarray}
\frac{F_K}{F_\pi}&=&1+\frac{4L_5^r}{F_\pi^2}(m_K^2-m_\pi^2)\,,\\[1mm]
\frac{m_K^2}{m_\pi^2}&=& 
\frac{\hat{m}+m_s}{2\hat{m}}\left[1-\frac{8(m_K^2-m_\pi^2)}{F_\pi^2}
(L_5^r-2L_8^r)\right]\label{l8fix}\,.
\end{eqnarray}

For the reason of brevity, in the analytic expressions of this section we 
omit scale-independent (finite) terms resulting from the loop corrections
which, nevertheless, are taken into account within the numerical analysis.
Note that in Eqs.~(\ref{fa1})--(\ref{l8fix}) we used $1/F_\pi$ and 
$2m_K^2/(\hat{m}+m_s)$ rather than the bare parameters $1/f$ and $r$. 
Formally, the difference represents higher order effects. Nevertheless, 
the appearance of $f$ or $r$ in Eqs.~(\ref{fa1})--(\ref{l8fix}) would induce 
a scale dependence which has no counterpart at the short-distance level
and will be absorbed at the next order in the chiral expansion.

The non-factorizable diagrams of Fig.~3 are calculated by introducing a
cutoff regulator $\Lambda_c$. Using the matching prescription of Fig.~1
we obtain
\begin{eqnarray}
i\langle \hspace{0.5mm}\pi^0\hspace{0.5mm}\pi^0\hspace{0.5mm}|Q_1|K^0
\rangle^{\NF}&=& 0\,,\\
i\langle \pi^+\pi^-|Q_1|K^0\rangle^{\NF}&=&\frac{X}{16\pi^2 F_\pi^2}\left(-3
\hspace{0.6mm}\Lambda_c^2+\frac{1}{4}\big(m_K^2+12m_\pi^2\big)
\ln\Lambda_c^2\right)\,,\hspace*{3mm}\label{nfa}\\
i\langle \hspace{0.5mm}\pi^0\hspace{0.5mm}\pi^0\hspace{0.5mm}|Q_2|K^0
\rangle^{\NF}&=& \frac{X}{16\pi^2 F_\pi^2}\left(\hspace{0.5mm}\frac{9}{2}
\Lambda_c^2+\frac{3}{4}\big(m_K^2-6m_\pi^2\big)\ln\Lambda_c^2\right)\,,\\
i\langle \pi^+\pi^-|Q_2|K^0\rangle^{\NF}&=& 
\frac{X}{16\pi^2 F_\pi^2}\left(\hspace{0.5mm}\frac{3}{2}\Lambda_c^2+
\big(m_K^2-\frac{3}{2}m_\pi^2\big)\ln\Lambda_c^2\right)\,,
\end{eqnarray}
\begin{eqnarray}
i\langle \hspace{0.5mm}\pi^0\hspace{0.5mm}\pi^0\hspace{0.5mm}|Q_6|K^0
\rangle^{\NF}&=& i\langle \pi^+\pi^-|Q_6|K^0\rangle^{\NF}\hspace{2mm}=
\hspace{2mm}\frac{X}{16 \pi^2 F_\pi^2}\frac{3}{4}
\left(\frac{2m_K^2}{\hat{m}+m_s}\right)^2 \ln\Lambda_c^2\,,\hspace*{5mm}\\ 
i\langle \hspace{0.5mm}\pi^0\hspace{0.5mm}\pi^0\hspace{0.5mm}|Q_8|K^0
\rangle^{\NF}&=&\frac{X}{16 \pi^2 F_\pi^2}\frac{3}{4}
\left(\frac{2m_K^2}{\hat{m}+m_s}\right)^2 \ln\Lambda_c^2\,,\\ 
i\langle \pi^+\pi^-|Q_8|K^0\rangle^{\NF}&=& -\frac{\sqrt{2}}{16\pi^2F_\pi}
\left(\frac{2m_K^2}{\hat{m}+m_s}\right)^2\frac{\alpha}{2}
\ln\Lambda_c^2\,.\label{nfo} 
\end{eqnarray}

The scale dependent terms in Eqs.~(\ref{nfa})--(\ref{nfo}) have to be
matched to the Wilson coefficient functions. To this end we use the
numerical values presented in the leading logarithm analysis of 
Ref.~\cite{Buch}.
%
\section{\boldmath $B_6^{(1/2)}$ \unboldmath and \boldmath $B_8^{(3/2)}$ 
\unboldmath}
To evaluate the hadronic matrix elements of $Q_6$ and $Q_8$ we pursued the 
following strategy. First, the non-factorizable contributions were calculated, 
in the isospin limit, from the diagrams of Fig.~3. In this part of the 
analysis the finite terms, which are systematically determined by the momentum 
prescription of Fig.~1, are completely taken into account. By using algebraic 
relations all integrals resulting from the diagrams of Fig.~3 can be reduced 
to three basic integrals which were calculated in the cutoff formalism.
Second, the finite terms arising from the factorizable loop diagrams of 
Fig.~2, as well as, from the renormalization of the wave functions, the 
masses and the low-energy couplings were estimated using dimensional 
regularization, as discussed at the end of Section 2. 

The values we obtain for the matrix elements of $Q_6$ and $Q_8$ are 
presented in Table~1, where we also specify the various contributions 
coming from the factorizable and the non-factorizable diagrams, respectively. 
In the numerical analysis we used the ratio $m_s/\hat{m}= 24.4\pm 1.5$ 
\cite{HL}. In the matrix elements we extracted the factor $R^2= 
[2m_K^2/(\hat{m}+m_s)]^2$, whose dependence on the (factorizable) 
scale is canceled exactly, for a general density-density operator,
by the diagonal evolution of the Wilson coefficients. 
%
\begin{table}[hbt]
\[ 
\begin{array}{|l||c|c|c|c|}\hline
&\lc=0.6\,\,\mbox{GeV}&\lc=0.7\,\,\mbox{GeV}&\lc=0.8\,\,\mbox{GeV}&\lc=0.9
\,\,\mbox{GeV} \\ 
\hline\hline 
\rule{0cm}{7mm}
i\langle Q_6\rangle_0^{\,\mbox{\footnotesize tree + F loops}} 
& -68.4-37.0i & -68.4-37.0i & -68.4-37.0i & -68.4-37.0i \\[1mm]
i\langle Q_6\rangle_0^{\,\mbox{\footnotesize NF loops}} 
& 29.8+37.0i & 34.6+37.0i & 39.0+37.0i & 42.9+37.0i
\\[1.5mm]
\hline\rule{0mm}{6.5mm}
|\langle Q_6\rangle_0|^{\,\mbox{\footnotesize total}} 
& 38.6 & 33.7 & 29.4 & 25.5 \\[1mm]
\hline\hline
\rule{0mm}{7mm}
i\langle Q_8\rangle_2^{\,\mbox{\footnotesize tree + F loops}} 
& 56.0-0.1i & 56.0-0.1i & 56.0-0.1i & 56.0-0.1i \\[1mm]
i\langle Q_8\rangle_2^{\,\mbox{\footnotesize NF loops}} 
& -20.7-11.5i & -24.8-11.5i & -28.8-11.5i & -32.8-11.5i \\[1.5mm]
\hline\rule{0mm}{6.5mm}
|\langle Q_8\rangle_2|^{\,\mbox{\footnotesize total}} 
& 37.2 & 33.2 & 29.5 & 25.9 \\[1.5mm]
\hline
\end{array}
\]
\caption{\footnotesize Hadronic matrix elements of $Q_6$ and $Q_8$
(in units of $R^2\hspace*{-0.7mm}\cdot\mbox{MeV}$), shown for various 
values of the cutoff $\lc$.}
\end{table}

In Table~2 we list the corresponding values for the $B_i$ factors which 
quantify the deviation of the hadronic matrix elements from the VSA results: 
\begin{equation}
B_6^{(1/2)}=|\langle Q_6 \rangle_0/\langle Q_6\rangle_0^{{\mbox 
{\scriptsize VSA}}}|\,,\hspace{1cm} 
B_8^{(3/2)}=|\langle Q_8 \rangle_2/\langle Q_8\rangle_2^{{\mbox 
{\scriptsize VSA}}}|\,.
\end{equation}
The VSA expressions for the matrix elements were taken from 
Eqs.~(XIX.16) and (XIX.24) of Ref.~\cite{BBL}. Numerically, they are
$|\langle Q_6 \rangle_0|=35.2\cdot R^2
\,\mbox{MeV}$ and $|\langle Q_8 \rangle_2|= 56.6\,\mbox{MeV}
\cdot [R^2-(0.389\,\mbox{GeV})^2]$. The second term in the expression for
$Q_8$ contributes at the $2\,\%$ level and has been neglected in Table~2.
%
\begin{table}[t]
\[ 
\begin{array}{|l||c|c|c|c|}\hline
&\lc=0.6\,\,\mbox{GeV}&\lc=0.7\,\,\mbox{GeV}&\lc=0.8\,\,\mbox{GeV}
&\lc=0.9\,\,\mbox{GeV} \\ \hline\hline
\rule[0mm]{0cm}{5.5mm}
B_6^{(1/2)}
& 1.10 & 0.96 & 0.84 & 0.72 \\[0.5mm]
B_8^{(3/2)}
& 0.66 & 0.59 & 0.52 & 0.46 \\[1.5mm]
\hline
\end{array}
\]
\caption{\footnotesize $B_6$ and $B_8$ factors for various values of 
$\Lambda_c$.} 
\end{table}
 
We discuss next the corrections to the matrix elements $\langle Q_6 
\rangle_0$ and $\langle Q_8 \rangle_2$. As already mentioned, the 
operator $Q_6$ is special because the ${\cal O}(p^0)$ tree level matrix 
element is zero due to the unitarity of the matrix $U$. Nevertheless the 
one-loop corrections to this matrix element must be computed. These 
corrections are of ${\cal O}(p^0/N_c)$ and are non-vanishing. 
We have shown in Eqs.~(\ref{nfa})--(\ref{nfo}) that the explicit 
calculation of the loops yields a cutoff dependence (i.e., a non-trivial 
scale dependence) from the non-factorizable diagrams which has to be 
matched to the short-distance contribution. We note that in the twofold 
expansion in $p^2$ and $p^0/N_c$ the contribution of the loops over the 
${\cal O}(p^0)$ matrix element must be treated at the same level as the 
leading non-vanishing tree contribution proportional to $L_5$. This is 
revealed by the large size of the non-factorizable ${\cal O}(p^0/N_c)$ 
corrections presented in Table~1. It is the sum of both, the factorizable 
and the non-factorizable contributions, which renders the numerical
values for $\langle Q_6\rangle_0$ close to the VSA value. 
The main effect of the loop corrections is to change the dependence 
of $\langle Q_6\rangle_0$ on $\lc$, from a flat behaviour at the tree level 
to the dependence presented in Tables~1 and 2. We note that at $\Lambda_c 
\simeq 700$ MeV the value for the matrix element of $Q_6$ is very close to 
the VSA result leading to $B_6\simeq 1$.

The $Q_8$ operator is not chirally suppressed, i.e., its ${\cal O}(p^0)$ 
tree level matrix element is non-zero. The one-loop corrections, even though 
suppressed by a factor $1/N_c$ with respect to the leading tree level, are 
found to be large and negative, leading to the small values for $B_8$ 
presented in Table~2. These large corrections persist in the octet limit 
[i.e, in the absence of the $\eta_0$, with $a=b=1$ and 
$m_\eta^2=(4m_K^2-m_\pi^2)/3\,]$. In comparison with $\langle Q_6\rangle_0$, 
the non-factorizable corrections to $\langle Q_8\rangle_2$ are less 
pronounced, as expected from the power counting scheme in $p^2$ and $1/N_c$. 
However, because the factorizable ${\cal O}(p^2)$ and ${\cal O}(p^0/N_c)$ 
corrections are small (and negative), the non-factorizable terms produce a 
significant reduction of $\langle Q_8\rangle_2$. The size of the higher 
order terms indicates that the leading-$N_c$ calculation or the closely 
related VSA are not sufficient for the matrix elements of the $Q_8$ operator.

It is interesting to compare our results with those of other calculations.
Refs.~\cite{JMS1} and \cite{EAP2} investigated $1/N_c$ corrections to the 
matrix elements of $Q_6$ and $Q_8$. This calculation considered the product 
of the two densities without the color singlet boson and the matching of 
short- and long-distance contributions was not explicit as in the present 
analysis. The numerical results showed also a tendency of reducing 
$\langle Q_8\rangle_2$ substantially, whereas $\langle Q_6\rangle_0$ was 
found to be enhanced compared to the VSA result. Calculations in lattice 
QCD obtain values for $B_6$ close to the VSA, $B_6^{(1/2)}(2\,\mbox{GeV})
=1.0\pm 0.2$ \cite{kil,sha} and 0.76(3)(5) \cite{peki}. Recent values 
reported for $B_8$ are $B_8^{(3/2)}(2\,\mbox{GeV})=0.81(3)(3)$ \cite{BGS}, 
0.77(4)(4) \cite{kgs}, and 1.03(3) \cite{Conti}. The chiral quark model 
\cite{Fras2} yields a range for $B_6$ which is above the VSA value, 
$B_6^{(1/2)}(0.8\,\mbox{GeV})=1.2-1.9$, and predicts a small reduction of 
the $B_8$ factor, $B_8^{(3/2)}(0.8\,\mbox{GeV})=0.91-0.94$. 
Although the scales used in the lattice calculations and the phenomenological 
approaches are different, the various results can be compared as the $B_6$ 
and $B_8$ parameters were shown in QCD to depend only very weakly on the 
renormalization scale for values above $1\,\mbox{GeV}$ \cite{BJL}. 

We note that our result for $B_6^{(1/2)}$ is in rough agreement with those of 
the various studies quoted above, whereas the value we obtain for
$B_8^{(3/2)}$ lies below the values reported previously. It is desirable 
to investigate whether this substantial reduction of $\langle Q_8\rangle_2$, 
which is due to the non-factorizable $1/N_c$ corrections to the leading term 
in the chiral expansion of $Q_8$, will be affected by higher order corrections.
This point is of great phenomenological interest because a less effective 
cancellation between the $Q_6$ and $Q_8$ operators, in the range obtained in 
the present analysis, will lead to a large value of $\varepsilon'/\varepsilon$ 
in the ball park of $\sim 10^{-3}$. This can be confirmed or disproved by the 
forthcoming experiments at CERN (NA48), Fermilab (E832) and Frascati (KLOE).
%
%
\section{$\Delta I = 1/2$ Rule}
Along the lines discussed in Section 2 we computed also the hadronic
matrix elements of the current-current operators $Q_1$ and $Q_2$, 
including the finite terms omitted in the analytic expressions of  Section 
3.\hspace*{1mm}\footnote{Note that due to current conservation no finite terms
arise from the factorizable loop diagrams of Fig.~2.} Subsequently, we carried
out the matching with the short-distance coefficient functions $z_i(\mu)$.
Our numerical result for the amplitude $A_0$ is depicted in Fig.~4. 
It shows an additional enhancement compared to the result of Ref.~\cite{BBG}
which renders the amplitude in good agreement with the observed value. The 
new contribution arises from the $Q_1$ and $Q_2$ operators. It is due 
to the modified matching prescription in the non-factorizable sector 
[except for approximately $20\,\%$ of it explained by the choice 
of the physical value $F_\pi$ in Eqs.(\ref{nfa})--(\ref{nfo})].\\[-2pt]

\noindent
\centerline{\epsfig{file=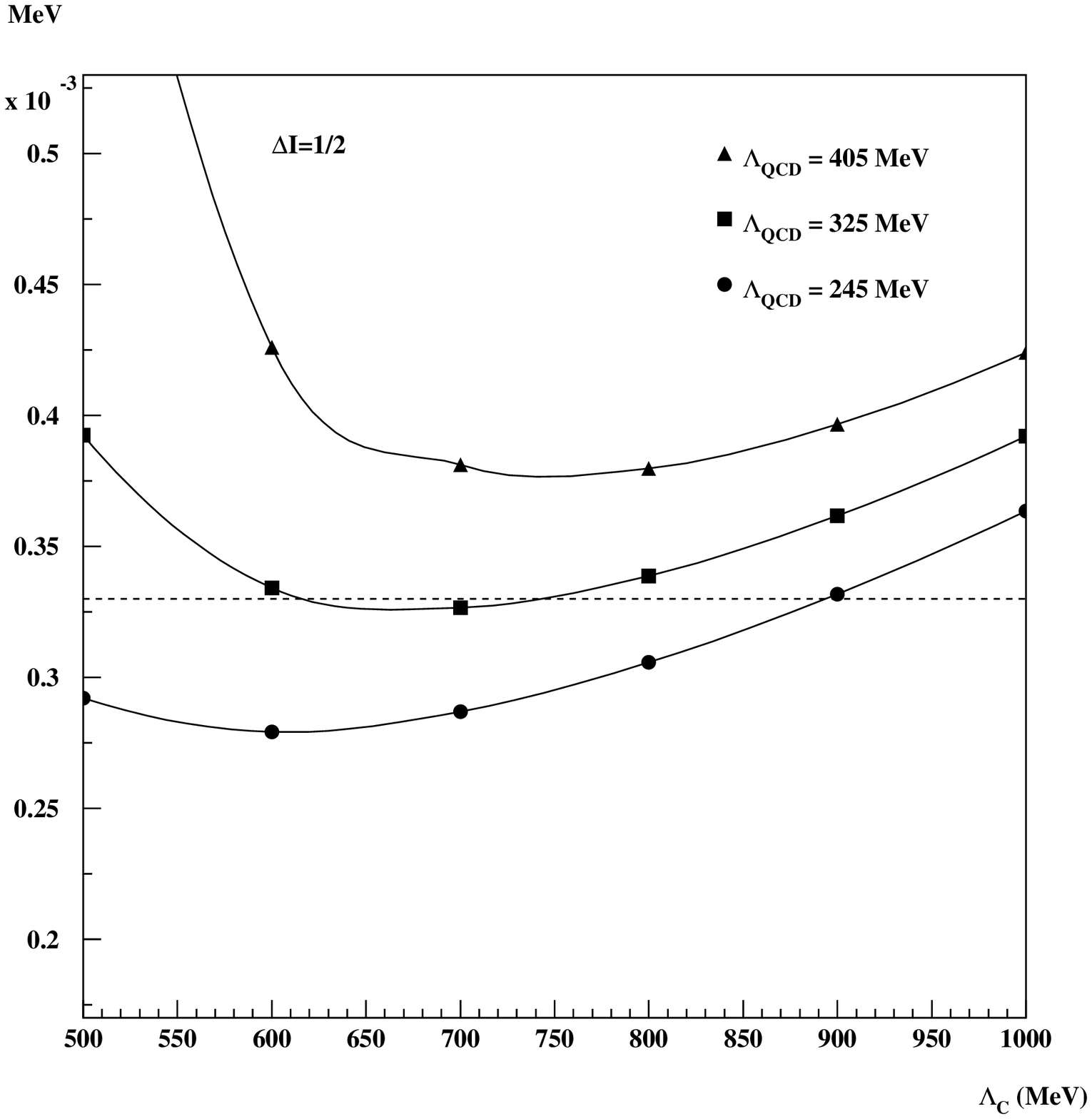,width=10cm}}\\[7pt]
\footnotesize
Fig.\ 4. $Re\,A_0$ in units of $10^{-3}$ MeV for $m_s(1\,$GeV)
$=150\,$MeV as a function of the matching scale $\Lambda_c$. The 
experimental value is represented by the dashed line.\\[3pt]
\normalsize

Our result is widely stable with respect to the matching scale. The
main uncertainty displayed in Fig.~4 originates from the dependence of the 
Wilson coefficients on $\Lambda_{\mbox{\tiny QCD}}$.
On the other hand, the $\Delta I=3/2$ amplitude depicted in Fig.~4 is highly 
unstable. The large uncertainty can be understood from the fact that it 
is obtained from the difference of two large amplitudes of approximately 
the same size [namely $A(K^0\rightarrow\pi^+\pi^-)$ and 
$A(K^0\rightarrow\pi^0\pi^0)$]. Consequently, the $\Delta I=3/2$ amplitude 
is not well reproduced except that, and this is an important point, it 
turns out to be sufficiently suppressed whatever the particular chosen 
scale is between $500\,$MeV and $1$\,GeV.

In conclusion, it is certainly premature to say that the dynamical
mechanism behind the $\Delta I = 1/2$ selection rule for $K\rightarrow\pi\pi$ 
decays is now completely understood since the $1/N_c$ expansion we use is 
only an approximate realization of non-perturbative QCD. In particular, 
vector mesons and higher resonances should be included in order to make it
possible to choose a higher value for the matching scale (it is probable 
that the vector mesons play an important role for $A_2$). Nevertheless we 
believe that the enhancement reported here is a further important indication 
that the $1/N_c$ approach can account for the bulk of the $\Delta I=1/2$ rule 
for $K\rightarrow\pi\pi$ decays.

\noindent
\centerline{\epsfig{file=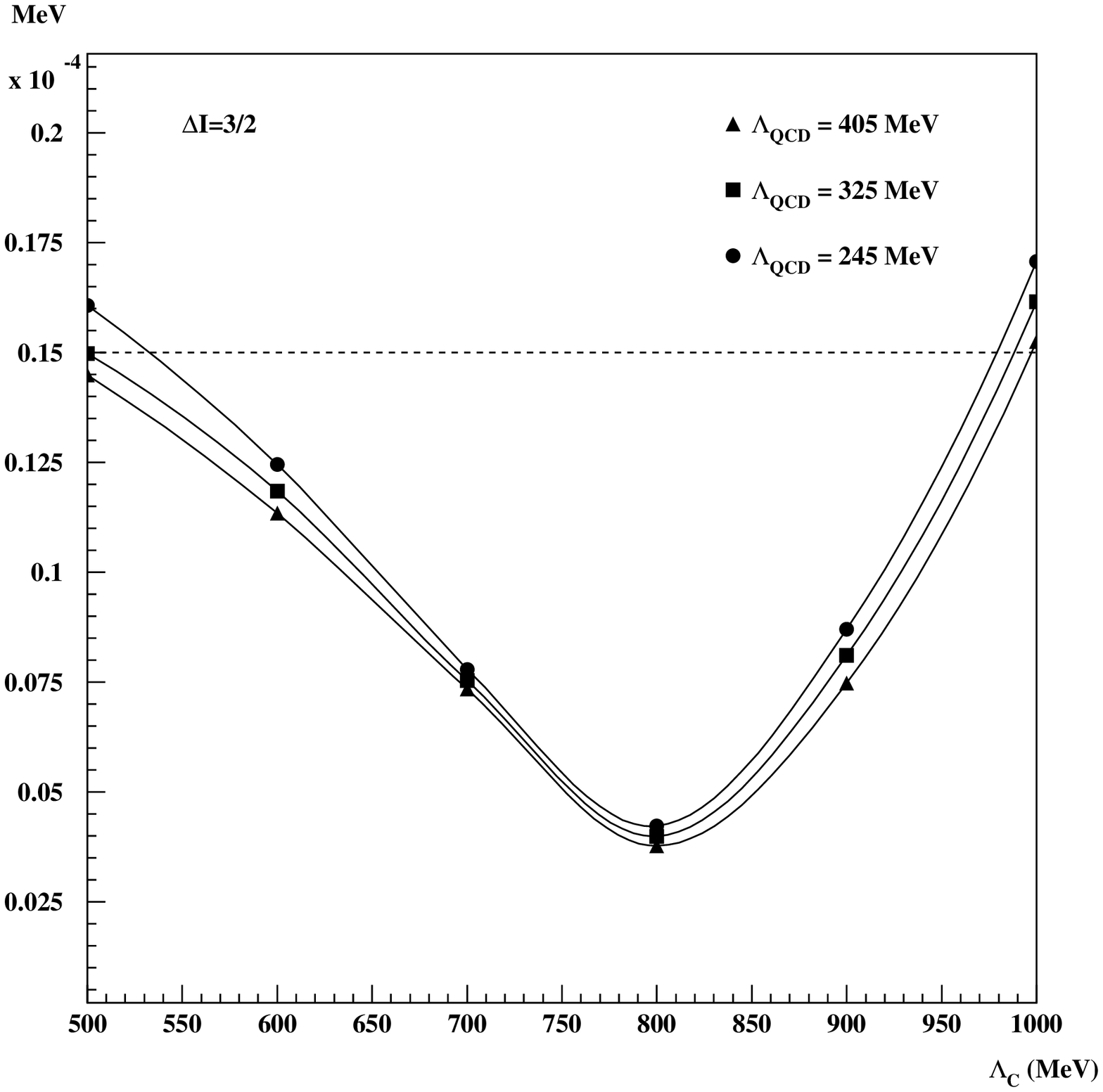,width=10cm}}\\[7pt]
\footnotesize
Fig.\ 5. $Re\,A_2$ in units of $10^{-3}$ MeV for $m_s(1\,$GeV)
$=150\,$MeV as a function of the matching scale $\Lambda_c$. The 
experimental value is represented by the dashed line.\\[24pt]
\normalsize

\noindent
\begin{center}{\large Acknowledgements}
\end{center}
This work has been done in collaboration with T. Hambye, E.A.~Paschos,
P. Soldan, and W.A.~Bardeen. We wish to thank J. Bijnens and J.-M. 
G\'erard for helpful comments. Financial support from the Bundesministerium 
f\"ur Bildung, Wissenschaft, Forschung und Technologie (BMBF), 057D093P(7), 
Bonn, FRG, and DFG Antrag PA-10-1 is gratefully acknowledged.
%
\vspace*{1.5cm}
\end{document}